\documentclass[letterpaper]{sfchem}
\usepackage{graphicx}

\newcommand{\too}{$\rightarrow$}
\newcommand{\HII}{H\,{\sc ii}}
\newcommand{\uchii}{UCH\,{\sc ii}}

\newcommand{\kms}{km\,s$^{-1}$}

\begin{document}

\title{Subarcsecond Imaging of Hot Cores with BIMA}

\author{Andy G. Gibb\inst{1} Friedrich Wyrowski\inst{2} \and Lee G. Mundy\inst{1} } 
  \institute{ Department of Astronomy, University of Maryland, College
  Park, MD 20742, USA
\and Max-Planck-Institut f\"ur Radioastronomie, Auf dem H\"ugel 69, 53121
  Bonn, Germany} 
\authorrunning{Gibb, Wyrowski, and Mundy}
\titlerunning{BIMA studies of hot cores}

\maketitle 

\begin{abstract}
\noindent
We present 1.4-mm BIMA observations with subarcsecond resolution of
the bright dust and molecular line emission from hot cores associated
with a sample of four ultracompact \HII\ regions: G9.62+0.19,
G10.47--0.03, G29.96--0.02, G31.41+0.31. Density power laws can
reproduce the observed continuum emission but break down on scales
smaller than 2000~AU. A total of 38 transitions from 18 species are
detected, with G10.47 and G31.41 showing the greatest number of lines.
In particular, these sources display emission from two
collisionally-excited transitions of methanol lying more than 950 K
above the ground state.  Outflows traced by H$_2$S emission
provide evidence for embedded exciting sources and the observed
morphology of molecular lines is consistent with internal heating of
the cores.


\end{abstract}

\section{Introduction: hot cores}
  
Hot cores are dense condensations of gas found in the vicinity of
ultracompact \HII\ (\uchii) regions. Their hot compact nature was
first revealed by ammonia observations (e.g. \cite{pauls83}), which
also provided the first clues to the different chemistry occurring in
hot cores (\cite{w87}). Compared with dark clouds, the abundances of
several species showed orders of magnitude enhancements which led to
the proposition that evaporation of grain mantles was important
(\cite{m86}; \cite{h87}). The inclusion of freeze-out onto dust grains
in chemical models yielded good agreement and gave the first clues to
the formation process of massive stars, namely that they must go
through a cold collapse phase of evolution (\cite{bcm88}).

Most studies of hot cores to date have been limited either spatially
or spectrally. Single-dish studies of the rich chemistry of hot cores
can yield valuable information on the species present and allow column
densities to be estimated (e.g. \cite{t91}; \cite{mghm96}).  However
in order to derive spatial information, high resolution observations
must be made.  The VLA has led the field here with subarcsecond
imaging of ammonia (\cite{c98}) which permits temperature and density
information to be derived on scales of a few thousand AU.  However, it
is millimetre/submillimetre interferometry which hold the key to
simultaneously probing the structure and chemistry of hot cores.
Furthermore, it is in the 1.3-mm window that current instruments can
make substantial gains with the wide variety of molecular lines that
are accessible and the fact that the continuum emission will be
dominated by emission from dust rather than free-free emission from
ionized gas in an \HII\ region. To date 1.3-mm observations have been
made towards the Orion hot core (\cite{blake96}) and W3(OH)
(\cite{fw97}, 1999), as well as two of our targets (G29.96 and
G31.41), albeit with lower resolution (\cite{maxia01}).

The proximity of hot cores to \HII\ regions led early studies to
suspect that hot cores may be heated by radiation from the
newly-formed OB stars within the \HII\ region. However, there is now
mounting evidence that at least some hot cores mark the sites of
massive star formation (e.g. \cite{hwh01}, \cite{t00}). In the next
section, results are presented of a new 1.4-mm BIMA study toward
\uchii/hot core regions, with the aim of separating \HII\ regions and
hot dust from the molecular cores. On the whole, our data support the
notion that our target hot cores contain embedded sources.

\section{BIMA observations}

The observations were made with the 9 elements of the BIMA array at
Hat Creek, California, equipped with 1.3-mm receivers. A sample of
four hot cores near ultracompact \HII\ regions (G9.62+0.19,
G10.47--0.03, G29.96--0.02, G31.41+0.31) was observed with angular
resolution varying from 0.4 (G10.47) to 2.8 arcsec (G9.62). The
correlator windows were set to cover a continuous bandwidth of 800 MHz
centred at 216.4 and 219.9 GHz with spectral resolution of 4.3
km\,s$^{-1}$. In addition, C$^{18}$O and H$_2$S were observed with
higher spectral resolution (1.1 km\,s$^{-1}$). The 1.4-mm continuum
images were made by summing the line-free channels in each band. All
four regions were classified as `line-rich' by \cite*{htmm}.

\section{Results I: 1.4-mm continuum}

In each case continuum emission was detected towards the hot core
positions at 1.4 mm, clearly resolved from the respective \uchii\ 
regions. Comparing our 1.4-mm images with VLA and our own BIMA 3-mm
data shows that the 1.4-mm emission is due to warm dust grains. Only
in G10.47 is the peak 1.4-mm continuum coincident with a \uchii\ 
region.  In G9.62 and G29.96 continuum emission is also detected
towards the \uchii\ region(s), which appears to be free-free emission
from the ionized gas, although in G29.96 there does appear to be some
warm dust within the shell of the \uchii\ region. The water masers
(\cite{hc96}) in each region are associated with the dust continuum
emission. Images comparing the dust continuum with the free-free
emission are given in \cite*{wgm02}.

The peak continuum flux ranges from 0.2 to 0.7 Jy per beam, equivalent
to brightness temperatures of up to 98 K (in a beam of 0.5 arcsec
towards G10.47). Thus the dust emission is warm. Masses of a few
hundred to a thousand solar masses are inferred from these data, which
translate to molecular hydrogen column densities of
$\sim10^{25}$\,cm$^{-2}$ towards the cores. Volume densities as high
as $10^8$\,cm$^{-3}$ are derived for G10.47 and G31.41.

The high angular resolution data enables us for the first time to
study the structure of the dust cores on scales of 0.1~pc and below.
Previous, single dish studies were limited in the modelling to larger
scales, hence probing only the outer envelope (see \cite{hfmtm}). We
have used the infrared radiative transfer models of \cite*{wc86}
to model the inner structure of the cores. Results of spherical dust
models containing embedded heating sources with luminosity of
$10^5$\,L$_\odot$ were `reobserved' with the spatial-frequency
coverage of the BIMA observations and then compared to the
observations directly in Fourier space. Density power laws (with
indices ranging from --1.5 to --2.0) can be fit to the data, but they
break down on scales of approximately 2000 AU. It is not yet clear
what happens within this radius.  Currently, we cannot distinguish
between, e.g., optically thick embedded disks (cf \cite{c98}) or, on
the other extreme, a dust free inner bubble.

\section{Results II: Spectral line data}

Emission was detected from 38 transitions from 18 species, although
not all lines and species were detected in each source.  G9.62 showed
the fewest lines ($\sim$12 for E, 15 for the hot core source G9.62F)
and the least number of high excitation transitions. In contrast,
G10.47 displayed 38 lines within the 1.6 GHz passband, a line density
of 25 lines per GHz. By way of comparison, single-dish studies of hot
cores have detected only half this quantity (\cite{mghm96}),
demonstrating not only the richness the spectrum of hot cores but also
how much potential information millimetre interferometry is capable of
yielding.

The lowest energy line detected was the 2\too 1 line of C$^{18}$O; the
highest energy is the 25(3)\too 24(4) $E$ transition of CH$_3$OH which
lies 986 K above the ground state. Note too that this methanol line is
a high-$J$ line and thus is probably collisionally excited in hot gas,
rather than pumped by infrared radiation from the warm dust.

In Figure~\ref{linedata} we present images of H$_2$S 2(2,0)\too
2(1,1), CH$_3$OH 7(1)\too 8(0)\,$E$ and HNCO 10(0,10)\too 9(0,9)
superimposed on the 1.4-mm continuum emission. 

\begin{figure*}
\centering 
\vspace{7in}
\caption{Top to bottom: G9.62, G10.47, G29.96 and G31.41. Contours are
  of molecular line emission superimposed on a linear colourscale
  representation of the 1.4-mm continuum. Filled squares mark the
  position of \uchii\ regions and crosses mark the positions of water
  masers. The filled ellipses represent the continuum beam dimensions;
  the outline ellipses are the spectral-line beams.
\label{linedata}}
\end{figure*}

\subsection{G9.62+0.19}

G9.62 contains up to 9 \HII\ regions (\cite{t00}; \cite{c94}), A to I.
Figure~\ref{linedata} marks the location of \HII\ regions D to G. The
hot core is G9.62F, while G9.62E also appears to be a deeply embedded
massive young stellar object. All the lines detected towards G9.62
peak at the position of F although most are also detected towards E.
SO and H$_2$S are the brightest lines towards both E and F.  The
linewidths are consistently higher towards the hot core source
(typically 8 \kms\ for F compared with 5 \kms\ for E), possibly due to
the presence of the outflow from F (\cite{hwh01}).

\subsection{G10.47--0.03}

G10.47 may be the most evolved of the sources here since there does
appear to be two \uchii\ regions embedded within the hot core. The
\uchii\ region marked B in Figure~\ref{linedata} splits into two (B1
and B2) at higher resolution (\cite{c98}). G10.47 has the brightest
and broadest emission lines in our sample. Since the continuum
brightness temperature is of order 100 K (see above) the presence of
40--50 K lines indicates gas temperatures in excess of 150 K. The
typical linewidth is $\sim$13 \kms.  The brightest line in the
passband is the 9(1,8)\too 9(1,9) transition of H$_2$CO (59 K).
Surprisingly the second brightest is the 24(4,20)\too 23(4,19)
transition of ethyl cyanide (52 K). Two very high excitation lines of
methanol (with upper energy levels exceeding 950 K above ground) were
detected towards G10.47.

As shown in Figure~\ref{linedata}, the emission from H$_2$S and
methanol is extended. Although no clear outflow is seen, the extension
of the emission is asymmetric and could be part of a bipolar outflow.
The ethyl cyanide (not shown) distribution resembles that of a shell,
supporting the outflow interpretation. On the whole, the
nitrogen-bearing species have a more compact distribution than
oxygen-bearing species. In particular, the HNCO lines show a more
compact appearance with increasing energy.

\subsection{G29.96--0.02}

G29.96 is a spectacular cometary \HII\ region (\cite{wc91}). Like
G34.26+0.15 (\cite{mghm96}; \cite{wm99}), the hot core lies close to
the tip of the cometary bow shape and is thus a prime candidate for
testing whether hot cores are internally heated by an embedded source
or externally heated by radiation from OB stars in the nearby \HII\ 
region. In our observations all line emission originates from the hot
core. No line emission is detected towards the \uchii\ region.
Furthermore, all the line emission peaks on the hot core itself with
no evidence for stratification of species by type or excitation. Most
notably, the HNCO lines (believed to be pumped by mid- and far-infared
radiation) all peak on the dust core, as do optically thinner species
such as H$_2 ^{13}$CO. Thus in G29.96 all evidence points towards the
hot core being internally heated, presumably by a luminous embedded
protostar.  The lack of line emission towards the \uchii\ region and
the presence of line emission to the west of the hot core further
supports the hypothesis that the cometary \HII\ region is a champagne
flow.

\subsection{G31.41+0.31}

G31.41 is the second richest source in our sample with 37 detected
lines. The two high-energy lines of methanol were detected towards
G31.41 as well. The lines are slightly narrower than those of G10.47
($\sim$ 10 \kms), although our spectral resolution is not sufficient
to distinguish such differences to high accuracy. The continuum
emission is slightly elongated in the same direction as the ammonia
(4,4) core detected by \cite*{c98} suggesting they are tracing the
same structure.

Intriguingly, and more clearly than in G10.47, the methanol and H$_2$S
shown in Figure~\ref{linedata} appear to trace a shell of emission
perhaps indicative of an outflow. Alternatively, the emission south of
the hot core (seen in the methanol image) may be a separate source since
it is evident in several other lines. The HNCO peaks on the hot core,
presumably at the position of the strongest infrared radiation field.

\subsection{Outflows from hot cores}

The higher-spectral resolution observations of H$_2$S revealed
non-Gaussian line wings towards all four targets. It is known that
G9.62F houses an embedded source which drives an outflow seen in
HCO$^+$ (\cite{hwh01}), although our H$_2$S does not show a clear
bipolar distribution. The most striking result is for G29.96 (shown in
Figure 2) in which spatially-bipolar red- and blue-shifted line wings
are seen centred on the hot core. The SO 6(5)\too 5(4) emission is
also extended in the direction of the H$_2$S lobes.  Furthermore, the
methanol masers of \cite*{mbc00} are aligned with this flow. This
result provides further evidence for the presence of luminous
protostellar sources embedded within hot cores.

\begin{figure*}
\centering
\includegraphics[width=0.8\linewidth]{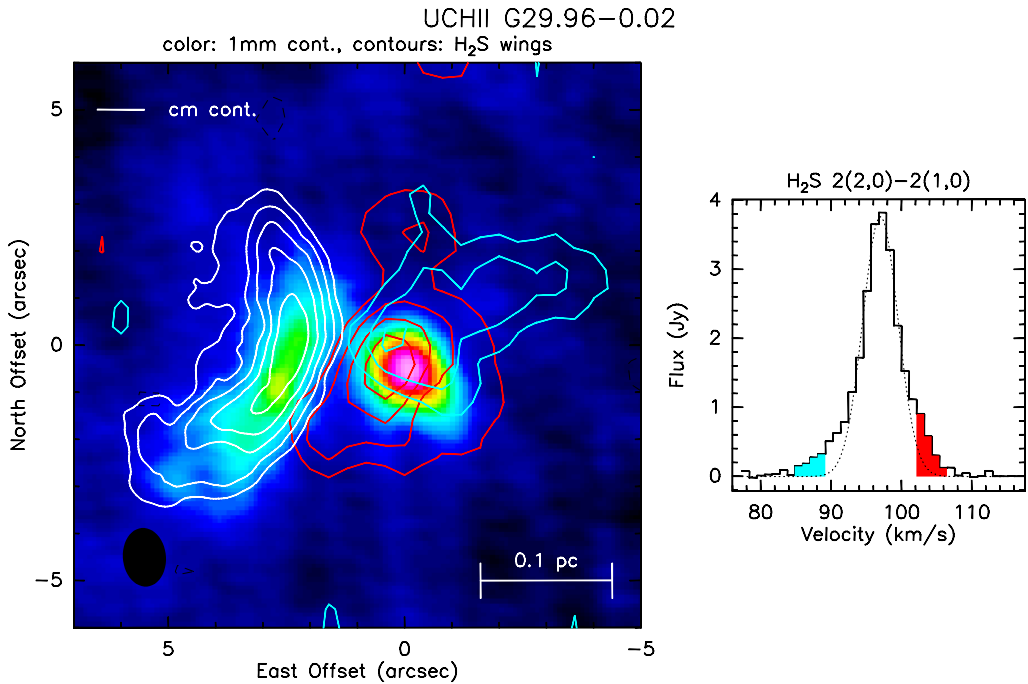}
\caption{H$_2$S outflow towards G29.96 (red and blue contours)
  compared with the 1.4-mm continuum (colourscale) and 1.3-cm
  free-free emission from \protect\cite*{c98} (white contours). The
  right panel shows the spectrum towards the hot core along with a
  Gaussian fit. The excess high-velocity wing emission is shaded for
  clarity and marks the velocity range for each lobe.}
\label{g29-h2s}
\end{figure*}
 
\section{Discussion: embedded sources in hot cores}

\begin{figure*}
\centering
\includegraphics[width=0.8\linewidth]{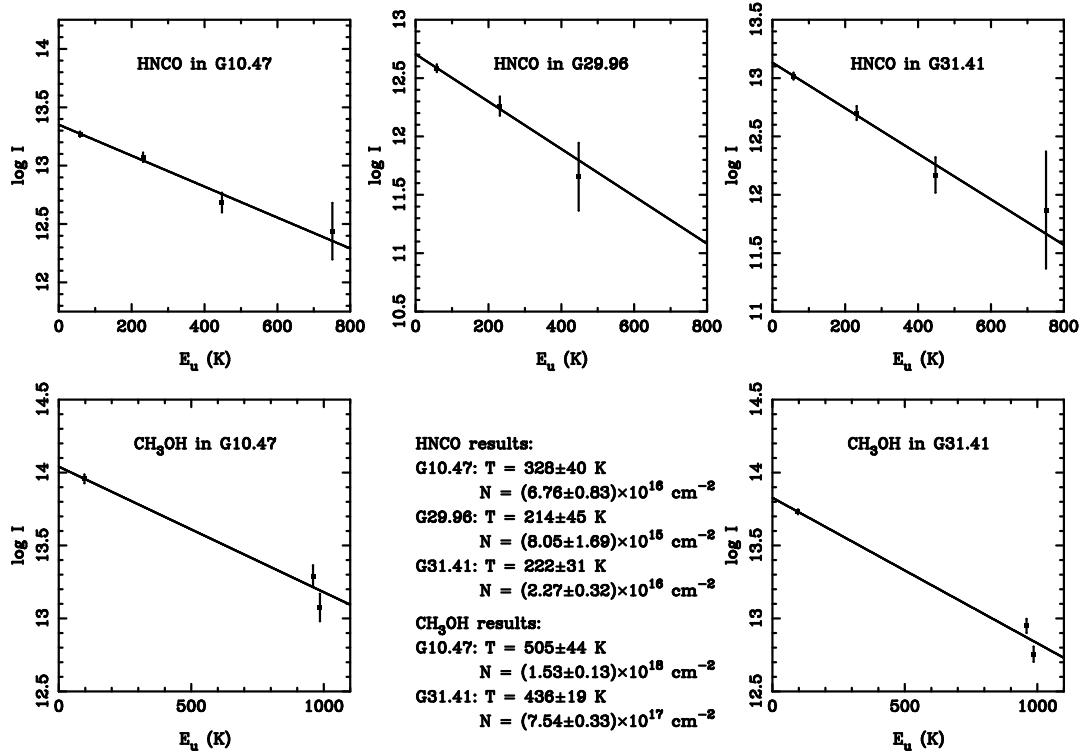}
\caption{Rotation diagrams for HNCO and methanol in G10.47, G29.96
  (HNCO only) and G31.41. The error bars represent the 1-$\sigma$
  uncertainties. The results of linear least-squares fits to these
  data are shown in the centre panel of the bottom row.
\label{rotdiag}}
\end{figure*}

In G10.47 and G31.41, three lines of methanol and four transitions of
HNCO were detected. Three transitions of HNCO were detected in G29.96.
We have constructed rotation diagrams (e.g. \cite{mghm96}) which are
shown in Figure~\ref{rotdiag}. These confirm the expectation from the
above results that G10.47 and G31.41 contain hotter cores than the
other sources in our sample with G10.47 being the hottest. Clearly
more data points are desirable, especially of methanol, in order to
constrain these fits further. The temperature of 505 K in G10.47 is
the highest yet determined from molecular line observations of hot
cores. However, it should be noted that the lowest energy line of
methanol used in this derivation may be optically thick, which would
give rise to an artificially high rotation temperature.

As pointed out above, the high energy lines of methanol are probably
excited by collisions as they come from levels with high $J$ (=25 and
24). Thus they reflect the actual gas kinetic temperature on radii of
1500 to 2000 AU from the central source. It is not certain whether
these temperatures represent those of the ambient gas or merely a
shocked component.

For the HNCO lines, it is likely that their excitation is via a strong
mid-infrared radiation field (\cite{cw86}). All the HNCO lines we have
detected are from the same $J$(upper) state but lie in different
$K_{-1}$ ladders (\cite{z00}).  Transitions between different
$K_{-1}$-levels are very rapid and require either very high densities
(exceeding 10$^9$\,cm$^{-3}$) or a mid- to far-infared radiation
field. Lower values of $K_{-1}$ are excited at longer wavelengths. The
detection of the $K_{-1}=4$ transition at 750 K above ground indicates
a significant 20-$\mu$m radiation field. \cite*{debuizer02} have
detected the hot core in G29.96 at 18 $\mu$m supporting the HNCO
detections.

\section{Conclusions}

We have used the BIMA millimetre array to observe a sample of four
\uchii\ regions at a wavelength of 1.4 mm with subarcsecond
resolution. In each case we detect compact dust cores spatially
distinct from the \uchii\ regions, showing that the hot cores mark the
location of a hot dense condensation of gas and dust. Dust emission is
also detected within the shell of the cometary \uchii\ region G29.96.
Modelling of the dust emission reveals density power laws with indices of
--1.5 to --2 beyond a radius of $\sim$2000 AU. It is tempting to
speculate that within this radius either no dust is present or the
geometry changes from spherical to axisymmetric, although further
observations with higher sensitivity on long baselines are required to
validate either of these options.

Line emission from 38 transitions of 18 molecular species has been
detected. G10.47 shows the greatest number of lines, with G31.41
coming a close second. In each of these sources two high-energy lines
of CH$_3$OH are observed with upper-level energies in excess of 950 K
above the ground state. The high $J$ number for these transitions
suggests that they are collisionally excited and are thus tracing
highly excited gas. The spectral line emission peaks towards the dust
cores with no strong evidence of excitation and/or chemical gradients,
which indicates that the hot cores are probably not externally heated.
Further evidence for internal heating comes from the observation of
spatially-resolved bipolar wing emission from H$_2$S in G29.96. Our
observations point towards these hot cores as housing massive,
luminous protostellar objects which have yet to develop a \uchii\
region. 

\begin{acknowledgements}
Research with BIMA is funded by NSF grant AST-9981289.
\end{acknowledgements}

\end{document}